\documentstyle[12pt]{article}

\begin{document}
\title{Search for flavor lepton number violation in slepton decays at LHC}
\author{N.V.Krasnikov \\Institute for Nuclear Research, Moscow 117312\\}

\date{November, 1996}
\maketitle
\begin{abstract}
We show that in supersymmetric models with explicit flavor lepton number 
violation due to soft supersymmetry breaking mass terms there could be 
detectable flavor lepton number violation in slepton decays. We  
estimate LHC discovery potential of the lepton flavor number 
violation in slepton decays.    

\end{abstract}

\begin{flushright}
preprint INR 927r/96
\end{flushright}     

\newpage

Supersymmetric electroweak models offer the simplest solution of the 
gauge hierarchy problem \cite{1}-\cite{4}. In real life supersymmetry has 
to be broken and the masses of superparticles have to be lighter than 
$O(1)$ Tev  \cite{4}. Supergravity gives natural explanation of the 
supersymmetry breaking, namely, an account of the supergravity breaking in 
hidden sector leads to soft supersymmetry breaking in observable sector 
\cite{4}. For the supersymmetric extension of the Weinberg-Salam model 
soft supersymmetry breaking terms usually consist of the gaugino mass 
terms, squark and slepton mass terms with the same mass at Planck scale and 
trilinear soft scalar terms proportional to the superpotential \cite{4}. 
For such "standard" supersymmetry breaking terms the lepton flavor number 
is conserved in supersymmetric extension of Weinberg-Salam model.
However, in general, squark and slepton soft supersymmetry 
breaking mass terms are not diagonal due to many reasons \cite{5}-\cite{15} 
(an account of stinglike or GUT interactions, nontrivial hidden sector, ..)
and  flavor lepton number is explicitly broken due to 
nondiagonal structure of slepton soft supersymmetry breaking mass terms.
As a consequence such models predict flavor lepton  number violation in  
$\mu$- and $\tau$-decays \cite{5}-\cite{13}. In our previous papers \cite{16}-
\cite{18} 
we proposed to look for flavor lepton number violation at LEP2 and NLC in 
slepton decays.

In this paper we investigate the LHC "discovery potential" of flavor 
lepton number violation in slepton decays. 
We find that at LHC it would be possible to discover lepton number violation 
in slepton decays for slepton masses up to 300 Gev provided that the mixing 
between sleptons is closed to the maximal one. 

In supersymmetric extensions of the Weinberg-Salam model supersymmetry is 
softly broken at some high energy scale $M_{GUT}$ by generic soft terms
\begin{eqnarray}
-L_{soft} = m_{3/2}(A^{u}_{ij}\tilde{u}_R^j\tilde{q}_L^iH_u + A^{d}_{ij}\tilde{d}^j_R\tilde{q}^i_LH_d + \nonumber\\
A^l_{ij}\tilde{e}_R\tilde{l}_LH_d +  h.c.) +  (m_q^2)_{ij}\tilde{q}_L^i(\tilde{q}^i_L)^+ + (m^2_u)_{ij}\tilde{u}^i_R \nonumber \\
(\tilde{u}^j_R)^+ +(m^2_d)_{ij}\tilde{d}^i_R(\tilde{d}^j_R)^+ + (m_l^2)_{ij}\tilde{l}^i_L(\tilde{l}^j_L)^+ +(m^2_e)_{ij}\tilde{e}^i_R \nonumber \\
(\tilde{e}^j_R)^+ + m^2_1H_uH_u^+ + m_2^2H_dH_d^+ + \nonumber \\
(Bm^2_{3/2}H_uH_d +\frac{1}{2}m_a(\lambda\lambda)_{a} + h.c.) \ ,  
\end{eqnarray}
where $i, j, a$ are summed over 1,2,3 and $\tilde{q}_{L}$, $\tilde{u}_{R}$,
$\tilde{d}_{R}$ denote the left- (right-)handed squarks, $\tilde{l}_{L}$,  
$\tilde{e}_{R}$ the left- (right-)handed sleptons and $H_u$, $H_d$ the two 
Higgs doublets; ${m}_a$ are the three gaugino masses of $SU(3)$, 
$SU(2)$ and $U(1)$ respectively. In most analysis the mass terms are supposed 
to be diagonal at $M_{GUT}$ scale and gaugino and trilinear mass terms are also 
assumed universal at $M_{GUT}$ scale. The renormalization group 
equations for soft parameters \cite{19} allow to connect high energy 
scale with observable electroweak scale. The standard consequence of such 
analysis is that righthanded sleptons $\tilde{e}_R$, $\tilde{\mu}_R$ and 
$\tilde{\tau}_{R}$ are the lightest sparticles among squarks and sleptons.
In the approximation when we neglect lepton Yukawa coupling constants they 
are degenerate in masses. An account of the electroweak symmetry breaking 
gives additional contribution to righthanded slepton square mass equal 
to the square mass of the corresponding lepton and besides an account 
of lepton Yukawa coupling constants in the superpotential leads to the 
additional contribution to righthanded slepton masses
\begin{equation}
\delta M^{2}_{sl} = O(\frac{h_{l}^{2}}{16{\pi}^2})M^{2}_{av}
ln(\frac{M_{GUT}}{M_{av}}) 
\end{equation}
Here $h_{l} $ is the lepton Yukawa coupling constant and $M_{av}$ 
is the average mass of sparticles. These effects lead to the splitting between 
the righthanded slepton masses of the order of
\begin{equation}
\frac{(m^2_{\tilde{\mu}_{R}} - m^2_{\tilde{e}_R})}{m^2_{\tilde{e}_R}} = 
O(10^{-5}) - O(10^{-3}) \ ,
\end{equation}
\begin{equation} 
\frac{(m^2_{\tilde{\tau}_R} - m^2_{\tilde{e}_R})}{m^2_{\tilde{e}_R}} = O(10^{-3}) - O(10^{-1}) 
\end{equation}
For nonzero value of trilinear parameter $A$ after electroweak symmetry 
breaking we have nonzero mixing between righthanded and lefthanded sleptons,
however the lefthanded and righthanded sleptons differ in masses (lefthanded 
sleptons are slightly heavier), so the mixing between righthanded and lefthanded
sleptons (for $\tilde{e}_R$ and $\tilde{\mu}_R$) is small and we 
shall neglect it. In our analysis we assume that the lightest stable particle 
is gaugino corresponding to $U(1)$ gauge group that is now more or less 
standard assumption \cite{20}. As it has been discussed in many papers 
\cite{5} - \cite{15} in general we can expect nonzero nondiagonal soft 
supersymmetry breaking terms in Lagrangian (1) that leads to additional 
contributions for flavor changing neutral currents and to flavor lepton number 
violation. From the nonobservation of $\mu \rightarrow e + \gamma $ decay
($Br(\mu \rightarrow  e + \gamma) \leq 5\cdot 10^{-11}$ \cite{22})  one can 
find that \cite{5,6}-\cite{19} 
\begin{equation}
\frac{(\Delta{m^{2}{_e\mu}})_{RR}}{M^{2}_{av}} \equiv 
(\delta_{e\mu})_{RR} \leq 2k\cdot 10^{-1}M^2_{av}/(1 Tev)^2,
\end{equation}
where $k = O(1)$. In our estimates we shall take $k = 1$. 
For $m_{\tilde{e}_{R}} =70 Gev$ we find that $(\delta_{e\mu})_{RR} 
\leq 10^{-3}$. Analogous bounds resulting from the nonobservation 
of $\tau \rightarrow e \gamma$ and $\tau \rightarrow \mu \gamma$ decays 
are not very stringent \cite{5,6}-\cite{23}.
  
The mass term for righthanded 
$\tilde{e}_{R}$ and $\tilde{\mu}_{R}$ sleptons has the form
\begin{equation}
-\delta{L} = m_{1}^{2}\tilde{e}_{R}^{+}\tilde{e}_{R} + 
m_{2}^{2}\tilde{\mu}_{R}^{+}\tilde{\mu}_{R} +
m^{2}_{12}(\tilde{e}_{R}^{+}\tilde{\mu}_{R} + 
\tilde{\mu}^{+}_{R}\tilde{e}_{R})
\end{equation}

After the diagonalization of the mass term (6) we find that the eigenstates of 
the mass term (6) are
\begin{equation}
\tilde{e}_{R}^{'} = \tilde{e}_{R}\cos(\phi) + \tilde{\mu}_{R}\sin({\phi}) \ , 
\end{equation}
\begin{equation} 
\tilde{\mu}_{R}^{'} = \tilde{\mu}_{R}\cos(\phi) - \tilde{e}_{R}^{'}\sin(\phi)
\end{equation}
with the masses
\begin{equation}
M^{2}_{1,2} = (1/2)[(m^2_1 + m^2_2) \pm ((m^2_1 - m^2_2)^2 + 
4(m^{2}_{12})^2)^{1/2}]
\end{equation}
which practically coincide for small values of $m^2_1 - m^2_2$ and 
$m_{12}^2$.
Here the mixing angle $\phi$ is determined by the formula
\begin{equation}
\tan(2\phi) = 2m^{2}_{12}(m^2_1 -m^2_2)^{-1}
\end{equation}
The crusial point is that even for small mixing parameter $m^{2}_{12} $
due to the smallness of the difference $m^{2}_{1} - m^2_{2}$ the mixing 
angle $\phi$ is in general not small (at present state of art it is 
impossible to calculate the mixing angle $\phi$ reliably).  
For the most probable case when the lightest stable superparticle is 
superpartner of the $U(1)$ gauge boson plus some small mixing with other 
gaugino and higgsino, the sleptons $\tilde{\mu}_R$, 
$\tilde{e}_R$ decay mainly into leptons $\mu_R$ and $e_R$ plus U(1) gaugino 
$\lambda$. The corresponding term in the Lagrangian responsible for 
slepton decays is
\begin{equation}
L_{1} = \frac{2g_{1}}{\sqrt{2}}(\bar{e}_{R}\lambda_{L}\tilde{e}_{R} + 
\bar{\mu}_{R}\lambda_{L}\tilde{\mu}_{R} +h.c.),\,  
\end{equation}
where $g_{1}^{2} \approx 0.13$. For the case when mixing is absent the 
decay width of the slepton into lepton and LSP is given by the formula
\begin{equation}
\Gamma = \frac{g^2_1}{8\pi}M_{sl}\Delta_{f}  
\approx 5\cdot10^{-3}M_{sl}\Delta_{f},\,
\end{equation}
\begin{equation}
\Delta_{f} = (1 - \frac{M^{2}_{LSP}}{M^{2}_{sl}})^{2} ,\, 
\end{equation}
where $M_{sl}$ and $M_{LSP}$ are the masses of slepton and the lightest 
superparticle (U(1)-gaugino) respectively. 
For the case of nonzero mixing we find that the Lagrangian (11)
in terms of slepton eigenstates reads
\begin{equation}
L_{1} = \frac{2g_{1}}{\sqrt{2}}[\bar{e}_{R}\lambda_{L}
(\tilde{e}_{R}^{'}\cos(\phi) - \tilde{\mu}_{R}^{'}\sin(\phi)) + 
\bar{\mu}_{R}\lambda_{L}(\tilde{\mu}^{'}_{R}\cos(\phi) + 
\tilde{e}_{R}^{'}\sin(\phi)) + h.c.]
\end{equation}

Let us now describe briefly the situation with the search for flavor 
lepton number violation in slepton decays at LEP2 and NLC.
At LEP2 and NLC in the neglection of slepton mixing   
$\tilde{\mu}_R$ and $\tilde{\tau}_R$ sleptons pair production occurs 
\cite{21} via annihilation graphs involving the photon and the $Z^{0}$ boson
and leads to the production of $\tilde{\mu}_R^+ \tilde{\mu}_R^-$ and 
$\tilde{\tau}_R^+ \tilde{\tau}_R^-$ pairs. For the production of 
righthanded selectrons in addition to the annihilation graphs we also have 
contributions from the t-channel exchange of the neutralino \cite{23} .
In the absence of mixing the cross sections can be represented in the 
form 
\begin{equation}
\sigma(e^+e^- \rightarrow \tilde{\mu}_{R}^{+}\tilde{\mu}_{R}^{-}) = 
kA^2 ,\,
\end{equation}
\begin{equation}
\sigma(e^+e^- \rightarrow \tilde{e}^{+}_{R}\tilde{e}^{-}_{R}) = 
k(A + B)^2 ,\,
\end{equation}
where $A$ is the amplitude of s-exchange, $B$ is the amplitude of t-exchange 
and $k$ is the normalization factor. The corresponding expressions for 
$A$, $B$ and $k$ are contained in \cite{23}. The amplitude $B$ is determined 
mainly by the exchange of the lightest gaugino and its account leads 
to the increase of selectron cross section by factor $k_{in} = (4-1.5)$. 
As it has been mentioned before we assume that righthanded sleptons are 
the lightest visible superparticles. 
So righthanded sleptons decay with 100 percent probability into leptons 
and LSP that leads to accoplanar events with missing transverse momentum. 
The perspectives for the detection of sleptons at LEP2 have been discussed 
in refs. \cite{23}-\cite{24} in the  assumption of flavor lepton number 
conservation. The main background 
at LEP2 energy comes from the $W$-boson decays into charged lepton and 
neutrino \cite{21}. For $\sqrt{s} = 190$ Gev the cross section of 
the $W^{+}W^{-}$ production is $\sigma_{tot}(W^{+}W^{-}) \approx 26pb$.
\cite{22}. For selectrons at $\sqrt{s} = 190$ Gev, selecting events with 
electron pairs with $p_{T,mis} \geq 10$ Gev and the accoplanarity angle 
$\theta_{ac} \geq 34^{\circ}$ \cite{21}, the only background effects 
left are from $WW \rightarrow e \nu e \nu$ and $e \nu \tau \nu$ where 
$\tau \rightarrow e \nu \nu$. For instance, for $M_{\tilde{e}_R} = 85$ Gev 
and $M_{LSP}=30 $ Gev one can find  that the accepted 
cross section is $\sigma_{ac}= 0.17 pb$ whereas the background 
cross section is $\sigma_{backgr} = 0.17 pb$ that allow to detect righthanded 
selectrons at the level of $5\sigma $ for the luminosity $150 pb^{-1}$ and 
at the level of $11\sigma $ for the luminosity $500 pb^{-1}$. For the detection 
 of righthanded smuons we have to look for  events with two accoplanar muons 
 however the cross section will be (4 - 1.5) smaller than in the selectron
case due to absence of t-channel diagram and the imposition of the cuts 
analogous to the cuts for selectron case allows to detect smuons for 
masses up to 80 Gev. Again here the main background comes from the $W$ decays 
into muons and neutrino. The imposition of more elaborated cuts allows to 
increase LEP2 righthanded smuon discovery potential up to 85 Gev on smuon 
mass \cite{23,24}.

Consider now the case of nonzero mixing $\sin{\phi} \neq 0$ between 
selectrons and smuons. In this case an account of t-exchange diagram 
leads to the following cross sections for the slepton pair production 
(compare to the formulae (15,16) ):
\begin{equation}
\sigma(e^+e^- \rightarrow \tilde{\mu}^{+}_{R}\tilde{\mu}^{-}_{R}) = 
k(A + B\sin^{2}(\phi))^2 ,\,
\end{equation}
\begin{equation}
\sigma(e^+e^- \rightarrow \tilde{e}^{+}_{R}\tilde{e}^-_R) = 
k(A + B\cos^{2}(\phi))^2 ,\,
\end{equation}
\begin{equation}
\sigma(e^+e^- \rightarrow \tilde{e}_{R}^{\pm}\tilde{\mu}^{\mp}_{R}) = 
kB^2\cos^{2}(\phi)\sin^{2}(\phi) 
\end{equation} 
Due to slepton mixing we have also lepton flavor number violation 
in slepton decays, namely:
\begin{equation}
\Gamma(\tilde{\mu}_{R} \rightarrow \mu + LSP) = \Gamma \cos^{2}(\phi),\,
\end{equation}
\begin{equation}
\Gamma(\tilde{\mu}_{R} \rightarrow e + LSP) = \Gamma \sin^{2}(\phi),\,
\end{equation}
\begin{equation}
\Gamma(\tilde{e}_{R} \rightarrow e + LSP) = \Gamma \cos^{2}(\phi),\,
\end{equation}
\begin{equation}
\Gamma(\tilde{e}_{R} \rightarrow \mu + LSP) = \Gamma\sin^{2}(\phi) 
\end{equation} 
Taking into account formulae (20-23) we find that 
\begin{eqnarray}
\sigma(e^+e^-  \rightarrow e^+e^- + LSP + LSP) =  k[(A + B\cos^{2}(\phi))^2\cos^{4}(\phi) \nonumber \\
+ (A + B\sin^{2}(\phi))^2\sin^{4}(\phi) + B^2\sin^{4}(2\phi)/8] ,\,
\end{eqnarray}
\begin{eqnarray}
\sigma(e^+e^- \rightarrow  {\mu}^+ {\mu}^{-} + LSP + LSP) = k[(A + B\cos^{2}(\phi))^2 \sin^{4}(\phi) \nonumber \\ 
+ (A + B\sin^{2}(\phi))^2\cos^{4}(\phi) + B^2\sin^{4}(2\phi)/8] ,\, 
\end{eqnarray}
\begin{eqnarray}
\sigma(e^+e^- \rightarrow {\mu}^{\pm} + {e}^{\mp} + LSP + LSP) = \frac{k\sin^{2}(2\phi)}{4}[(A + B\cos^{2}(\phi))^2 \nonumber \\ 
+ (A + B\sin^{2}(\phi))^2 + B^2(cos^{4}(\phi) + \sin^{4}(\phi))] 
\end{eqnarray}

It should be noted that formulae (24-26) are valid only in the approximation 
of narrow decay width of sleptons
\begin{equation}
2\Gamma m_{\tilde{e}_{R}} \leq |m^{2}_{\tilde{\mu}_{R}} - 
m^{2}_{\tilde{\mu}_{R}}|
\end{equation}
For the case when the inequality (27) does not hold the effects due to the 
finite decay width are important and decrease the cross section with violation 
of flavor lepton number. The cross section for the reaction 
 $ e^{+}e^{-} \rightarrow e^{+}{\mu}^{-} + LSP + 
LSP $  is 
proportional to 
\begin{equation}
\sigma \sim 
\sin^{2}(\phi)\cos^{2}(\phi)
\int |D(p_{1},m_{\tilde{e}},\Gamma)D(p_{2},
m_{\tilde{e}},\Gamma) - 
D(p_{1},m_{\tilde{\mu}},\Gamma)D(p_{2},m_{\tilde{\mu}},\Gamma)|^{2}
dp_{1}^{2}dp_2^2 ,\,
\end{equation}
where 
\begin{equation}
D(p,m,\Gamma) = \frac{1}{p^2 - m^2 -i\Gamma m}
\end{equation}
and $\Gamma_{\tilde{e}} \approx \Gamma_{\tilde{\mu}} = \Gamma$. 
The approximation (24-26) corresponds to the neglection of the interference 
terms in (28) and it is valid if the inequality (27) takes place. For 
smaller slepton masses difference an account of the interference terms in (28) 
is very important \cite{18,25}. The integral (28) is approximately equal to
\begin{equation}
\sigma \sim \sin^{2}(\phi)\cos^{2}(\phi) 
\frac{2\pi^2}{b^2}I_{dil},
\end{equation}
\begin{equation} 
I_{dil} = 
(1 - \frac{b^2(b^2 - \frac{a^2}{4})}{(b^2 + 
\frac{a^2}{4})^2 }) ,
\end{equation}
where $a = m^2_{\tilde{e}_{R}} - m^2_{\tilde{\mu}_R}$ , $b = \Gamma \cdot (
\frac{m_{\tilde{e}_R} +  m_{\tilde{\mu}_R}}{2})$. Here $I_{dil}$ determines 
the effect of destructive interference. 
 An account of the 
interference effects leads to the decrease of the cross section (26) by 
factors 1, 0.82, 0.52, 0.17 for $|m^{2}_{\tilde{e}} - m^{2}_{\tilde{\mu}}| 
= 2\Gamma m_{\tilde{e}}$ $1.5\Gamma_{\tilde{e}}$, $\Gamma m_{\tilde{e}}$, 
$0.5\Gamma m_{\tilde{e}}$ respectively. 

To be concrete consider the case of maximal selectron-smuon mixing
($m_1 = m_2$ in formula (6)). For this case bound (5) resulting from the 
absence of the decay $\mu \rightarrow e \gamma$ reads 
$\frac{|m^2_{12}|}{m_1^2} \leq  10^{-3}$ $(m_{\tilde{e}} = 70 Gev)$ ; 
$4.5\cdot10^{-3}$ $(m_{\tilde{e}} = 150 Gev)$ ; $8\cdot 10^{-3}$ 
$(m_{\tilde{e}} = 200 Gev)$ ; $10^{-2}$ $(m_{\tilde{e}} = 225 Gev)$. 
From the requirement (27) of the absence of the destructive interference 
for the cross section with flavor lepton number violation we find that
\begin{equation}
\frac{|m^2_{12}|}{m_1^2} \geq 5\cdot 10^{-3} {\Delta}_f
\end{equation}
So we see that for LEP2 energies it is possible to improve $\mu \rightarrow 
e \gamma$ bound only for the case when LSP mass is closed to slepton mass. 
For instance, for $m_{LSP} = 0.95m_{\tilde{e}}$, $0.9m_{\tilde{e}}$, 
$0.8m_{\tilde{e}}$, $0.7m_{\tilde{e}}$, 
$0.6m_{\tilde{e}}$, $0.5m_{\tilde{e}}$ 
$\Delta_{f}$ is equal to 0.01, 0.036, 0.13, 0.26, 0.41, 0.56 respectively. 
For $m_{\tilde{e}} = 70 Gev$ the destructive interference is 
not essential for $m_{LSP} \geq 0.74m_{LSP}$. For the Next 
Linear Collider energies and for $m_{\tilde{e}} \geq 100 Gev$ the 
$\mu \rightarrow e \gamma$ bound (5) is not so stringent and it is 
possible to improve it even for the case of relatively small LSP 
masses when $\Delta_{f} \approx 1$.  

The perspectives for the detection of righthanded  
sleptons at NLC (for the case of zero 
slepton mixing) have been discussed in ref.\cite{26}. The standard assumption 
of ref.\cite{26} is that righthanded  
sleptons are the NLSP, therefore the  only possible 
decay mode is $\tilde{l} \rightarrow l + LSP$. One possible set of selection 
criteria is the following:

1. $\theta_{acop} \geq 65^{\circ}$.

2. $p_{T,mis} \geq 25$ Gev.

3. The polar angle of one of the leptons should be larger than $44^{\circ}$, 
the other $26^{\circ}$.

4. $(m_{ll} - m_{Z})^2 \geq 100$ $Gev^2$.

5. $E_{l^{\pm}} \geq 150$ Gev.

For $\sqrt{s} = 500$ Gev and for integrated luminosity 20 $fb^{-1}$ , a 
$5\sigma$ signal can be found up to 225 Gev provided the difference between 
lepton and LSP is greater than 25 Gev \cite{26}. Following ref.\cite{26} we 
have analyzed the perspective for the detection of nonzero slepton mixing at 
NLC. In short, we have found that for $M_{LSP}=100$ Gev it is possible to 
discover selectron-smuon mixing at the $5\sigma$ level for $M_{sl}=150 $ 
Gev provided that $\sin{2\phi} \geq 0.28$. For $M_{sl}= 200$ Gev it is 
possible to detect mixing for $\sin{2\phi} \geq 0.44$ and  $M_{sl}=225$ 
Gev corresponds to the limiting case of maximal mixing ($\sin{2\phi} =1$) 
discovery.      

It should be noted that up to now we restricted ourselves to the case 
of smuon selectron mixing and have neglected stau mixing with selectron and 
smuon. Moreover for the case of stau-smuon or stau-selectron mixings 
bounds from the absence of $\tau \rightarrow \mu \gamma$ and 
$\tau \rightarrow e \gamma$ decays for $m_{\tilde{\tau}} \geq 70 Gev$ 
are not stringent \cite{5,6}-\cite{21}.  
For instance, for the case of stau-smuon mixing in formulae 
(24-26) we have to put $B = 0$ ( only s-exchange graphs contribute to the 
cross sections) and in final states we expect as a result of mixing 
${\tau}^{\pm}{\mu}^{\mp}$ accoplanar pairs. The best way to detect 
$\tau$ lepton is through hadronic final states, since 
$Br(\tau \rightarrow hadrons + {\nu}_{\tau})$ = 0.74. Again, in this case the 
main background comes from W-decays into $(\tau)^{\pm}(\mu)^{\mp} + \nu + \nu$
in the reaction $e^+e^- \rightarrow W^+W^-$. The imposition of some cuts 
\cite{23,24} decreases W-background to $0.07pb$ that allows to detect 
stau-smuon mixing for slepton masses up to 70 Gev. We have found that 
for $m_{sl} = 50 $ Gev it would be possible to detect mixing angle 
$\sin(2{\phi}_{\tau\mu})$ bigger than 0.70. Other detectable consequence of 
big stau-smuon mixing is the decrease of accoplanar ${\mu}^+ {\mu}^- $ 
events compared to the case of zero mixing. For instance, for the case of 
maximal mixing $\sin(2{\phi}_{\tau\mu}) = 1$ the suppression factor is 2.
In general we can't exclude also big mixing between all three 
righthanded sleptons. 

Consider now the possibility to discover lepton number violation 
in slepton decays at LHC. The possibility to discover sleptons at LHC 
have been discussed in refs.\cite{27}-\cite{29}. The main mechanism 
of slepton production at LHC is the Drell-Yan mechanism, 
so formulae (24-26) with $B =0$ are valid in our case. We shall use the 
results of ref.\cite{29} where concrete estimates have been made for 
CMS detector. 
To be concrete we consider two points of the ref.\cite{29}:

Point A: $m(\tilde{l}_{L}) = 314$ Gev, $m(\tilde{l}_{R}) = 192$ Gev, 
$m(\tilde{\nu}) =308$ Gev, $m(\tilde{\chi}_1^0) =181$ Gev, 
$m(\tilde{\chi}_2^0) = 358$ Gev, $m_(\tilde{g}) = 1036$ Gev, 
$m(\tilde{q}) =905$ Gev, $\tan(\beta) = 2$, $sign(\mu) = -$.

Point B: $m(\tilde{l}_L) =112$ Gev, $m(\tilde{l}_{R}) = 98$ Gev, 
$m(\tilde{\nu}) = 93$ Gev, $m(\tilde{\chi}^0_1) = 39$ Gev, 
$m(\tilde{\chi}^0_2) = 87$ Gev, $m(\tilde{g}) = 254$ Gev, $m(\tilde{q}) = 
234$ Gev, $\tan(\beta) =2$ , $sign(\mu) = -$ .

For point A the following cuts have been used:
$p^l_T \geq 50$ Gev, $Isol \leq 0.1$, $|\eta| \leq 2.5$, $E_{T}^{miss} 
\geq 120 $ Gev, $\Delta\phi(E_{T}^{miss}, ll) \geq 150^{o}$, jet veto - 
no jets 
with $E_{T}^{jet} \geq 30$ Gev in $|\eta| \leq 4.5$, Z-mass cut - $M_{Z} 
\pm 5$ Gev excluded, $\Delta\phi(l^+l^-) \leq 130^o$. With such cuts for 
the total luminosity $L_{t} = 10^{5} pb^{-1}$ 91 events $e^{+}e^{-} + 
{\mu}^{+}{\mu}^{-}$ resulting from slepton decays have been 
found. The standard  WS model background comes from 
$WW$, $t\bar{t}$ , $Wt\bar{b}$, $WZ$, $\bar{\tau}\tau$ and gives 
105 events. No SUSY 
background have been found. The significance for the slepton discovery at 
point A is 6.5. Using these results it is trivial to estimate the 
perspective for the discovery of flavour violation in slepton decays. Consider
 the most optimistic case of maximal slepton mixings 
(for both righthanded and lefthanded sleptons) and neglect the 
effects of destructive interference. 
For the case of maximal selectron-smuon mixing  the number of signal events 
coming from slepton decays is
 $N_{sig}(e^+e^-) = N_{sig}(\mu^{+}\mu^{-}) = N_{sig}(\mu^{\pm}e^{\mp}) = 
23 $. The number of background  events is $N_{back}(e^+e^-) = 
N_{back}(\mu^{+}\mu^{-}) = N_{back}(e^{\pm}\mu{\mp}) = 53$. The significance 
$S = \frac{Sleptons}{\sqrt{Background + Sleptons}}$ is 5.2 for all dilepton 
modes. For the case of maximal 
smuon-selectron mixing we have the same number of $e^{+}e^{-}, \mu^{+}\mu^{-}, 
e^{\pm} \mu^{\mp}$  signal events, whereas in the case of the mixing absence 
we don't have $e^{\pm} \mu^{\mp}$ events.  For the case of the maximal 
stau-smuon mixing we expect 23 $\mu^{+}\mu^{-}$ signal events and 46 
$e^+e^-$ signal events and 2 $\mu^{\pm}e^{\mp}$ signal events whereas the 
background is the same as for the case of maximal smuon-selrctron mixing. 
The significance is: 4.6($e^+e^-$ mode), 2.6($\mu^{+}\mu^{-}$ mode), 
5.2($e^+e^- + \mu^+\mu^-$ - mode). The case 
of selectron-stau mixing is the similar to the case of smuon-stau mixing the 
single difference consists in the replacement of $e \rightarrow \mu$ , 
$\mu \rightarrow e$. For the case of maximal selectron-smuon-stau mixing 
we expect  46 $e^+e^{-} + \mu^{+}\mu^{-} + e^{\pm}\mu^{\mp}$ signal events and 
the significance is 2.8. 

For the point B the cuts are similar to the point A, except $p^{l}_{T} 
\geq 20$ Gev, $E^{miss}_{T} \geq 50$ Gev, $\Delta\phi(E_{T}^{miss},ll) \geq 
160^{o}$ For the total luminosity $L_{tot} = 10^{4}pb^{-1}$ the number of 
$e^+e^- + \mu^+\mu^-$  events resulting from direct slepton production has 
been found to be 323. The number of background events have 
been estimated equal to 989(standard model background) + 108(SUSY background)=
1092. The significance is 8.6. Our analysis 
for the point B is similar to the corresponding analysis for the point A. 
For the case of maximal selectron-smuon mixing we have found that the 
significance for all delepton modes is 6.4. For the case of the 
maximal smuon-stau mixing the significance for $e^+e^- + \mu^{+}\mu^{-}$ 
mode is 6.6 . The same significance is for the case of the  maximal 
selectron-stau mixing. For the case of maximal selectron-smuon-stau mixing 
the significance for $e^+e^- + \mu^+\mu^{-} + e^{\pm}\mu^{\mp}$ mode is
3.0. For the total luminosity $L_{tot} = 10^{5}pb^{-1}$ the significance is 
increased by factor $\approx 3.1$. 
It is interesting to mention that at LHC the main mechanism of slepton pair 
production is the Drell-Yan mechanism and as a consequence for equal smuon and 
selectron masses the corresponding cross sections and the number of 
$e^{+}e^{-}$ and $\mu^{+}\mu^{-}$ signal events coincide. The corresponding 
cross sections depend rather strongly on slepton masses. If smuon and 
selectron masses differ by 20 percent the corresponding cross sections and 
as a consequence the number of $e^{+}e^{-}$ and $\mu^{+}\mu^{-}$ signal 
events will differ by factor $\approx 2$ that as it has been demonstrated on 
the example of points A and B is detectable at LHC. However the effect of 20 
percent smuon and selectron mass difference will imitate the effect of 
selectron-stau or smuon-stau mixings. So the situation could be rather 
complicated. At any rate by the neasurement of the differnce in 
$\mu^{+}\mu^{-}$ and $e^{+}e^{-}$ events it would be possible to measure 
the difference of smuon and selectron masses with the accuracy 
$\approx 20 percent$ that is very important because in MSSM  smuon and 
selectron masses practically coincide for both righthanded and lefthanded 
sleptons.
             
Let us formulate the main result of this paper: in supersymmetric extension of 
standard Weinberg-Salam model there could be soft supersymmetry breaking terms
responsible for flavor lepton number violation and slepton mixing. 
At LHC it would be possible to discover flavor lepton number violation in 
slepton decays for sleptons lighter than 300 Gev provided that the mixing 
among sleptons is closed to the maximal one. For the case of nonequal 
smuon and selectron masses the number of $e^+e^-$ and $\mu^+\mu^-$ 
events will be 
different that imitate the effect of stau-smuon or stau-selectron mivings. 
At any rate the observation (or nonobservation) of the $(\mu^{+}\mu^{-}$ - 
$e^{+}e^{-})$ difference allows to conclude that smuon and selectron masses 
differ(coincide) at least with the accuracy  20 percent or to make 
conclusion about the discovery of 
slepton mixing. Unfortunately it is rather difficult to distinguish between 
these two possibilities. For the case of nonzero smuon-selectron mixing the 
number of $\mu^+\mu^-$ and $e^+e^-$ events is predicted to be the same and 
moreover for the case of maximal smuon-selectron mixing the number of 
$\mu^{+}e^{-}$ and $\mu^{-}e^{+}$ events coincide with the number of 
$\mu^{+}\mu^{-}$ and $e^+e^-$ events. Of course, it is clear that at 
NLC or $\mu^+\mu^-$  collider the perspectives for the flavor lepton number 
violation discovery are the most promising but unfortunately now it is 
too far from reality.       
        
I thank CERN TH Department for the hospitality during my stay at CERN where 
this paper has been finished. I am indebted to the collaborators of the INR 
theoretical department for discussions and critical comments. I am indebted 
to Lali Rurua for very useful discussions.

\end{document}